\documentclass[%
 aip,
 amsmath,amssymb,
 reprint,%
]{revtex4-1}

\usepackage{graphicx}% Include figure files
\usepackage{dcolumn}% Align table columns on decimal point
\usepackage{bm}% bold math
\usepackage[utf8]{inputenc}
\usepackage[T1]{fontenc}
\usepackage{mathptmx}
\usepackage{etoolbox}
\usepackage{color}
\usepackage{romannum}

\makeatletter
\def\@email#1#2{%
 \endgroup
 \patchcmd{\titleblock@produce}
  {\frontmatter@RRAPformat}
  {\frontmatter@RRAPformat{\produce@RRAP{*#1\href{mailto:#2}{#2}}}\frontmatter@RRAPformat}
  {}{}
}%
\makeatother

\begin{document}

\title{Efficient Hartree-Fock Exchange Algorithm with Coulomb Range Separation and Long-Range Density Fitting}

\author{Qiming Sun}
\email{osirpt.sun@gmail.com}
\affiliation{Quantum Engine LLC, Washington 98516, US}

\date{\today}

\begin{abstract}
Separating the Coulomb potential into short-range and
long-range components enables the use of different electron repulsion integral
algorithms for each component.
The short-range part can be efficiently computed using the analytical algorithm due to the
locality in both Gaussian-type orbital basis and the short-range Coulomb potentials.
The integrals for the long-range Coulomb potential can be approximated with the
density fitting method.
A very small auxiliary basis is sufficient for the density fitting
method to accurately approximate the long-range integrals.
This feature significantly reduces the computational efforts associated with the
$N^4$ scaling in density fitting algorithms.
For large molecules, the range separation and long-range density fitting method
outperforms the conventional analytical integral evaluation scheme employed in
Hartree-Fock calculations and provides more than twice the overall performance.
Additionally, this method yields higher
accuracy compared to regular density fitting methods.
The error in the Hartree-Fock energy
can be easily reduced to 0.1 $\mu E_h$ per atom, which is significantly more
accurate than the typical error of 10 $\mu E_h$ per atom observed in regular
density fitting methods.
\end{abstract}

\maketitle

\section{Introduction}

A well-known bottleneck in Hartree-Fock and Kohn-Sham density functional theory
is the computation of the two-electron repulsion integrals (ERIs)
\begin{equation}
  (\mu\nu|\kappa\lambda)
  = \iint \mu(\mathbf{r}_1) \nu(\mathbf{r}_1)\frac{1}{r_{12}}
  \kappa(\mathbf{r}_2) \lambda(\mathbf{r}_2) d^3\mathbf{r}_1 d^3\mathbf{r}_2
  \label{eq:eri}
\end{equation}
for four atomic orbitals basis functions $\mu(\mathbf{r})$, $\nu(\mathbf{r})$,
$\kappa(\mathbf{r})$, $\lambda(\mathbf{r})$.
To improve the computation speed of ERIs, the density fitting (DF) or
resolution of identity methods were developed to approximate
ERIs \cite{Whitten1973,Dunlap1979,Vahtras1993,Eichkorn1995,Dunlap2000}.
%,
%\begin{gather}
%  (\mu\nu|\kappa\lambda)
%  \approx \sum_{PQ} (\mu\nu|\phi_P) [M^{-1}]_{PQ} (\phi_Q|\kappa\lambda),
%  \\
%  M_{PQ} = (\phi_P|\phi_Q),
%\end{gather}
%where $\phi_P$ is the function of auxiliary basis set.
This technique has been widely and successfully applied in quantum chemistry simulations for both molecules
and crystals\cite{Velde1991,Jaffe1996,Carsky2012,Burow2009,Feyereisen1993,Komornicki1993,Manby2001,Jung2005,Hollman2014,Sun2017,Ye2021,Lee2022}.
Despite the great success that DF has achieved, it still faces
several challenges in terms of computation costs and accuracy.

The DF method exhibits notable performance advantages when computing Hartree-Fock
exchange (HFX) in small and medium-size systems.
However, these advantages become less pronounced for larger molecules with more
than 10,000 basis functions. 
When applying the DF method to calculate HFX, the integral problem is transformed into a tensor
contraction problem with a scaling of $N^4$.
The performance advantage of the DF method can be attributed to the
computational benefits provided by high-performance tensor contraction software,
such as MKL, OpenBLAS, BLIS, Cutlass, as well as hardware such as GPUs.
For sufficiently large systems, the conventional analytical ERI computation for
HFX may actually be faster than DF HFX.
Due to the locality of Gaussian-type orbital (GTO) basis and the sparsity of the
GTO integral tensor, the conventional method has an asymptotic scaling between $N^1$
and $N^2$.
Besides the scaling wall, another challenge that restricts the application of
the DF method to large systems is the issue of I/O and storage.
The DF integral tensor is a three index tensor, with two indices representing
the orbital basis and one index representing the auxiliary basis.
Typically, the number of auxiliary basis functions is more than three times the
number of orbital basis functions.
In systems with 10,000 atomic orbitals, the size of a DF tensor can quickly reach tens or even hundreds of terabytes.
Even considering the sparsity of the DF tensor, the compressed tensor can still
occupy a significant amount of storage.

The effectiveness of the DF approximation heavily relies on the quality of the auxiliary basis.
To find a balance between computational cost and accuracy, the auxiliary basis
must be optimized in advance for the specific orbital basis and computation tasks.
Different computation tasks,
such as the computation of Coulomb energy (denoted as JFIT on basis exchange\cite{Pritchard2019}),
the Coulomb and exchange terms (JKFIT), the post-HF methods, and relativistic
methods, have different requirements on the auxiliary basis.
If an inappropriate auxiliary basis is used,
the DF approximation may result in reduced accuracy, or increased computational costs, or both.
To address this issue, various approaches
have been proposed to automate the construction of auxiliary
basis sets\cite{Yang2007,Bostroem2009,Aquilante2009,Kallay2014,Stoychev2017,Lehtola2021,Lehtola2023,DeChavez2023}.
The automatic basis generation procedure can simplify the selection of auxiliary basis
sets in calculations, although it may slightly increases computational costs.

In our previous work\cite{Sun2023a}, we developed the algorithm of ERI computation in
crystal systems with the separation of short-range (SR) and long-range (LR) in Coulomb potential
\begin{equation}
  \frac{1}{r_{12}}
  = \frac{\mathrm{erfc}(\omega r_{12})}{r_{12}} + \frac{\mathrm{erf}(\omega r_{12})}{r_{12}}.
\end{equation}
SR ERIs are computed analytically in real space, while LR ERIs are computed numerically on
reciprocal grids.
In this decomposition treatment, we observed that the rank of the LR ERI
tensor is significantly lower than that of the full-range ERIs.
Depending on the value of the parameter $\omega$ in the range-separation
operators, the rank of LR ERIs can be reduced to 1/100 of the rank of full-range ERIs.
This feature can be leveraged to construct low-rank DF integral tensors, which
effectively solves the problems of auxiliary basis dependence, I/O pressure and performance penalty in regular DF methods.
In this work, we will test the combination of range-separation treatment and the
long-range density fitting (LRDF) ERIs for the computation of Coulomb and HFX
terms in molecules.

\section{\label{sec:theory}Theory}
\subsection{Auxiliary Gaussian basis for long-range density fitting}
The range separation of the Coulomb potential enables us to decompose the ERI
tensor into SR and LR components,
\begin{equation}
  (\mu\nu|\kappa\lambda)
  = (\mu\nu|\kappa\lambda)_\text{SR} + (\mu\nu|\kappa\lambda)_\text{LR}.
\end{equation}
The SR ERI tensor contains details of the ERI structure. Its rank is similar
to that of the full-range ERI tensor.
The LR ERI tensor is a low-rank tensor.
It is efficient to use DF technique to construct this tensor with auxiliary
basis $\phi_P(\mathbf{r})$
\begin{equation}
  (\mu\nu|\kappa\lambda)_\text{LR}
  = \sum_{PQ} (\mu\nu|\phi_P)_\text{LR} [M^{-1}]_{PQ} (\kappa\lambda|\phi_Q)_\text{LR}.
\end{equation}
where
\begin{gather}
  (\mu\nu|\phi_P)_{LR}
  = \iint\mu(\mathbf{r}_1) \nu(\mathbf{r}_1)
  \frac{\mathrm{erf}(\omega r_{12})}{r_{12}}
  \phi_P(\mathbf{r}_2) d^3\mathbf{r}_1 d^3\mathbf{r}_2.
  \\
  M_{PQ}
  = \iint\phi_P(\mathbf{r}_1)
  \frac{\mathrm{erf}(\omega r_{12})}{r_{12}}
  \phi_Q(\mathbf{r}_2) d^3\mathbf{r}_1 d^3\mathbf{r}_2.
\end{gather}
By performing pivoted Cholesky decomposition (CD)\cite{Lehtola2019} or
eigenvalue decomposition (ED) on the metric matrix $\mathbf{M}$,
\begin{equation}
  \mathbf{M} = \mathbf{U} \lambda \mathbf{U}^\dagger,
\end{equation}
we can obtain the long-range density fitting (LRDF) tensor $\mathbf{L}$ with reduced auxiliary dimension
\begin{equation}
  L_{\mu\nu,P} = \sum_Q \lambda_Q^{-1/2} U_{PQ} (\mu\nu|\phi_Q)_\text{LR}.
\end{equation}
The transformed coefficients $\mathbf{U}$ can be truncated to form a short and
wide matrix, with fewer rows than columns. It only needs to retain the rows that
correspond to non-zero values of $\lambda$.

A straightforward choice for the auxiliary basis in LRDF would be the regular DF
auxiliary basis that was
previously optimized for the full-range Coulomb potential. However, these types of
auxiliary basis sets are typically over-complete and too large, compared to the
rank of the LR ERI tensor.
Since the LR Coulomb potential screens the
interactions between nearby density distributions, the auxiliary basis for LRDF
only needs to ensure an accurate description of the effective potential
experienced by the remote electron density distributions, which are
the products of two orbital basis functions $\mu(\mathbf{r})\nu(\mathbf{r})$.
This requirement is much lower than the requirements for the auxiliary
basis in regular DF methods.
Therefore, a smaller auxiliary basis with enough flexibility to represent
potentials in the multipole expansion is sufficient for the decomposition of LR ERIs.
The numerical tests in Section \ref{sec:results} suggest that primitive Gaussian basis
functions up to angular momentum $l=2$ with exponent of 1.0 on each atom are
enough to represent the LR potential.
In practice, using only $s$-type Gaussian functions
for Hydrogen atoms and Gaussian functions with $l=0..2$ for the remaining atoms can
achieve an accuracy of micro-Hartree per atom.

\subsection{Computing LR ERIs with PW basis}
The LR ERI can be Fourier transformed to an integral in reciprocal space
\begin{equation}
  (\mu\nu|\kappa\lambda)_\text{LR}
  = \frac{1}{(2\pi)^3}\int \frac{4\pi}{G^2}e^{-\frac{G^2}{4\omega^2}}
  \rho_{\mu\nu}(\mathbf{G}) \rho_{\lambda\kappa}(\mathbf{G})^* d\mathbf{G}^3.
  \label{eq:lr:gspace}
\end{equation}
Eq. \eqref{eq:lr:gspace} is essentially a one-particle Gaussian integrals.
The integrand
$\rho_{\mu\nu}(\mathbf{G})$ can be obtained through a
Fourier transformation of the orbital product $\rho_{\mu\nu}(\mathbf{r})$
\begin{equation}
  \rho_{\mu\nu}(\mathbf{G})
  = \int e^{-i\mathbf{G}\cdot\mathbf{r}} \rho_{\mu\nu}(\mathbf{r}) d\mathbf{r}^3
  = e^{-\frac{G^2}{4\alpha_{\mu\nu}}} e^{-i\mathbf{G}\cdot \mathbf{R}_{\mu\nu}}
  P_{\mu\nu}(\mathbf{G}).
  \label{eq:ft:rhoij}
\end{equation}
Here, $P_{\mu\nu}(\mathbf{G})$ is a polynomial associated with the Fourier
transform of the polynomial part in Gaussian product theorem\cite{Besalu2011}.
By changing the coordinates system to spherical coordinates,
\begin{equation}
  d^3\mathbf{G} = G^2 \sin\theta d\theta d\varphi,
\end{equation}
we can transform the integral \eqref{eq:lr:gspace} to
\begin{equation}
  \frac{1}{4\pi^2}\int e^{-\frac{G^2}{4\omega^2}}\rho_{\mu\nu}(G,\theta,\varphi)
  \rho_{\lambda\kappa}(G,\theta,\varphi)^* \sin\theta dG d\theta d\varphi.
  \label{eq:eri:spherical:G}
\end{equation}

Assuming that we have already evaluated the angular part of the integrals,
the remaining integral for the radial part $G$ can be expressed formally as
\begin{gather}
  \int_0^\infty e^{-\frac{G^2}{4\theta}} S(G) P(G) dG,
  \label{eq:integral:G}
  \\
  \theta = (\alpha_{\mu\nu}^{-1} + \alpha_{\kappa\lambda}^{-1} + \omega^{-2})^{-1},
  \\
  \alpha_{\mu\nu} = \alpha_\mu + \alpha_\nu.
\end{gather}
$S(G)$ is a factor associated to the coordinates of the four Gaussian
orbital basis functions.
\begin{gather}
  S(G) \sim \int e^{-i \mathbf{G}|\mathbf{R}_{\mu\nu} - \mathbf{R}_{kl}|} d\theta d\varphi,
  \label{eq:structure:factor}
  \\
  \mathbf{R}_{\mu\nu} = \frac{\alpha_\mu \mathbf{R}_\mu + \alpha_\nu
  \mathbf{R}_\nu}{\alpha_{\mu\nu}}.
\end{gather}
The other terms in the radial part are collected in the polynomial $P(G)$.

When the value of $\omega$ is selected such that $\omega^2 \ll \alpha_{ij}$ and
$\omega^2 \ll \alpha_{kl}$, the exponential term in the integral
\eqref{eq:integral:G} can be approximated as $e^{-u^2}$, where $u = \frac{G}{2\omega}$.
The integral can then be numerically evaluated using
Gauss-Hermite quadrature with respect to the weight function $e^{-u^2}$. 
To generate the grids $G_i$ and weights $w(G)_i$ for the radial part integral,
we use the roots and weights of Gauss-Hermite quadrature $u_i$ and $w(u)_i$ in the following manner
\begin{gather}
  G_i = 2\omega u_i, \\
  w(G)_i = 2\omega w(u)_i.
\end{gather}

The angular part of the integral involves a two-dimensional integration for
$\theta$ and $\varphi$,
\begin{equation}
  \int_{0}^\pi \sin\theta d\theta \int_{0}^{2\pi} d\varphi.
\end{equation}
It is convenient to change the variable to $t = \cos \theta$, and transform the
limits of angular integration to
\begin{equation}
  \int_{-1}^{1} dt \int_{0}^{2\pi} d\varphi.
\end{equation}
The numerical integration for the angular part can
then be performed using grids and weights on a spherical surface.
One possible choice for the angular grids is the Lebedev scheme, which is
widely used in numerical integration for density functional theory.
It is also possible to generate the angular grids by applying Gauss-Legendre quadratures
for $t$ and $\varphi$ separately.
Gauss-Legendre quadratures are preferred in this case.
They are more flexible in the number of grids that can be used and they are often
more accurate than the Lebedev grids when using the same number of angular grids.

Combining the Gauss-Hermite quadratures for the radial part and Gauss-Legendre
quadratures for the angular part, we obtain the integration grids (planewave
basis) and weights in reciprocal space
\begin{align*}
  G_x &= G_i \sin\arccos t_j \cos\varphi_k, \\
  G_y &= G_i \sin\arccos t_j \sin\varphi_k, \\
  G_z &= G_i \cos\arccos t_j, \\
  w &= \frac{1}{2\pi^2} G_i^2 w(G)_i w(t)_j w(\varphi)_k,
\end{align*}
where $w(t)_j$ and $w(\varphi)_k$ are the weights of the Gauss-Legendre quadratures for $t$ and $\varphi$, respectively.
It should be noted that the structure factor $S(G)$ in \eqref{eq:structure:factor} can result in a polynomial of
infinite order. It is necessary to have high resolution in both the radial and
angular parts to fully account for this effect.
The number of integration grids required for accurate integration is system
dependent.

By using these integration grids and weights, the LR integral \eqref{eq:lr:gspace} can
be computed
\begin{equation}
  (\mu\nu|\kappa\lambda)_\text{LR}
  = \frac{1}{2\pi^2}\sum_{n} \frac{w_n}{|\mathbf{G}_n|^2}
  \rho_{\mu\nu}(\mathbf{G}_n) \rho_{\lambda\kappa}(\mathbf{G}_n)^*
\end{equation}

The numerical integration in reciprocal space have several advantages.
\begin{itemize}
\item PW basis does not respond to the displacement of atoms.
It makes the computation of nuclear gradients and hessian with PW basis much simpler
than the procedure with Gaussian auxiliary fitting basis\cite{Bostroem2013}.
\item This method does not require I/O access. The Fourier transform
\eqref{eq:ft:rhoij} can be performed on the fly. The runtime generation is more
efficient than reading data from disk.
\item Parallelization is straightforward.
The LR potential in reciprocal space has a diagonal structure, which allows us
to isolate the FT of orbital products for each grid point.
The computation of FT on different grids can be easily distributed to different computing nodes.
\end{itemize}

\subsection{SCF algorithm}
In the SCF program, the SR and LR parts of the Coulomb and exchange matrices are computed separately.
The SR part can be incrementally constructed during SCF iterations using the
AO-direct algorithm\cite{Almloef1982}.
Integral screening is a crucial factor for the performance of SR integrals\cite{Izmaylov2006,Thompson2017}.
We employed the three hierarchical screening prescription developed in
our earlier works\cite{Sun2023,Sun2023a}. Given integral screening threshold $\varepsilon$
and density matrix $\gamma$,
the SR integral $(\mu\nu|\kappa\lambda)_\text{SR}$ can be discarded if it satisfies
\begin{equation}
  \begin{aligned}
  \max(\gamma_{\mu\nu},\gamma_{\kappa\lambda},
  \gamma_{\mu\kappa}, \gamma_{\mu\lambda},
  \gamma_{\nu\kappa}, \gamma_{\nu\lambda}) & \\
  \cdot \min( \sqrt{(\mu\nu|\mu\nu)_\text{SR} (\kappa\lambda|\kappa\lambda)_\text{SR}}, &\\
    \sqrt{(\mu\mu|\kappa\kappa)_\text{SR} (\nu\nu|\lambda\lambda)_\text{SR}}, &\\
    \sqrt{(\mu\mu|\lambda\lambda)_\text{SR} (\nu\nu|\kappa\kappa)_\text{SR}}, &\\
    \frac{Q_{\mu\nu}Q_{\kappa\lambda}e^{-\theta R^2}}{R^2} &)
  < \varepsilon.
  \end{aligned}
\end{equation}
In the above inequality, $Q_{\mu\nu}$ is a factor that behaves like the overlap
between two primitive Gaussian functions, as defined in reference \onlinecite{Sun2023}.
$R$ is the distance between two orbital pairs
\begin{equation}
R = |\mathbf{R}_{\mu\nu}-\mathbf{R}_{\kappa\lambda}|.
\end{equation}
For the LR part HFX, we can leverage the orbital coefficients
$\mathbf{C}$ and occupation numbers $\Lambda$ to reduce computational costs
\begin{gather}
  L_{\mu i,P} = \sum_\nu L_{\mu i,P} C_{\nu i}, \\
  K_{\mu\kappa} = \sum_{i P} \Lambda_i L_{\mu i,P} L_{\kappa i,P}.
  \label{eq:df:hfx}
\end{gather}
The computational cost of LR HFX can be further reduced if one employs
the occ-RI-K algorithm \cite{Manzer2015}.

\section{\label{sec:results}Numerical tests and discussions}
We have implemented the LRDF approach based on the PySCF
software package v2.3.0\cite{Sun2020}.
The SR integrals in current work are computed using Libcint v6.0\cite{Sun2015}.
Previous versions of the Libcint library can also be used, but the performance
is significantly slower for $s$-type and $p$-type GTOs.

In the first set of tests, we focus on the impact of various factors on the LRDF ERIs,
including the auxiliary basis, the range-separation parameter $\omega$,
the threshold of ED, and the numerical integration grids in reciprocal space.
To assess the accuracy of LRDF ERIs,
we compute the HF energy using the conventional analytical 4-center ERIs, and
compare this energy
with the one-shot HF energies of the LRDF methods in terms of various parameter settings.
The one-shot HF energy is computed using the density matrix obtained from the converged HF with the analytical ERIs.

To simulate the possible scenarios in real applications, we employ two systems in
these tests: a water cluster consisting of 12 water molecules, and a
linear molecule made up of 6 glycine monomers (Gly$_6$).
They can represent two extreme cases of basis distributions in real space.
In the water cluster, orbital basis and auxiliary basis have significant overlap,
making it relatively easy to expand the orbital product $\rho_{\mu\nu}$ with the auxiliary basis.
In the linear molecule Gly$_6$, it is more challenging to use auxiliary basis to
describe the orbital products as the auxiliary basis is sparsely distributed in space.
With the same parameter settings, linear molecules should generally produce larger errors.
Results are collected in Table \ref{tab:water12:auxbasis} - \ref{tab:gly6:auxbasis}.

In the second set of tests, we utilize the parameters determined from the
previous tests to evaluate the SCF performance of the LRDF method on large
molecules. The test systems consist of a water cluster with 128
water molecules and a linear molecule with 30 glycine monomers (Gly$_{30}$).
Orbital basis set cc-pVDZ is employed throughout all tests.
The results of these tests are presented in Table \ref{tab:scf}.

\begin{table}
  \centering
  \caption{The overall errors in HF energy of (H$_2$O)$_{12}$ with auxiliary Gaussian basis.
  The reference HF energy is -912.40220404.}
  \label{tab:water12:auxbasis}
  \begin{tabular}{llrrlrlrlrrrr}
\hline
\multicolumn{2}{l}{Auxiliary Basis} & $N_{aux}$ & \multicolumn{2}{c}{$\omega=0.1$}
     & \multicolumn{2}{c}{$\omega=0.15$} & \multicolumn{2}{c}{$\omega=0.2$} \\
O & H & & $N_{ED}$ & $\Delta E/\mu E_h$ & $N_{ED}$ & $\Delta E/\mu E_h$ & $N_{ED}$ & $\Delta E/\mu E_h$ \\
\hline
\multicolumn{5}{l}{ED threshold $10^{-4}$} \\
 $s$      &$s$                      & 36  & 30  & 91.66 & 36  & 2848. & 36  & 18380. \\
%$2s$     &$s$                      & 48  & 35  & -176. & 47  & 1206. & 48  & 9721.  \\
%$2s$     &$2s$                     & 72  & 40  & -98.0 & 56  & 675.9 & 71  & 6539.  \\
 $sp$     &$s$                      & 72  & 36  & -120. & 56  & -9.97 & 70  & 899.5  \\
 $sp$     &$sp$                     & 144 & 46  & -4.59 & 80  & -40.8 & 107 & -88.8  \\
%$2sp$    &$s$                      & 84  & 38  & -348. & 63  & -15.1 & 80  & 441.8  \\
 $spd$    &$s$                      & 132 & 39  & -124. & 72  & -71.5 & 104 & -96.0  \\
 $spd$    &$spd$                    & 324 & 49  & 0.31  & 96  & -38.0 & 152 & -57.4  \\
%$2spd$   &$s$                      & 144 & 40  & -233. & 75  & -81.7 & 109 & -69.0  \\
%$2s2p2d$ &$s$                      & 240 & 66  & -66.9 & 104 & -346. & 141 & 8.154  \\
%$2s2p2d$ &$spd$                    & 432 & 68  & -51.9 & 113 & -106. & 182 & -10.4  \\
%$2s2p2d$ &$2s2p2d$                 & 648 & 89  & -12.2 & 156 & -50.6 & 225 & -196.  \\
\multicolumn{2}{l}{weigend}         & 852 & 64  & -37.7 & 113 & -72.4 & 169 & -13.1  \\
\multicolumn{2}{l}{cc-pvdz-jkfit}   & 1392& 68  & -36.3 & 124 & -25.6 & 193 & -7.63  \\
\hline
\multicolumn{5}{l}{ED threshold $10^{-5}$} \\
 $s$      &$s$                      & 36  & 35  & 110.8 & 36  & 2848. & 36  & 18380. \\
%$2s$     &$s$                      & 48  & 42  & 31.50 & 48  & 1186. & 48  & 9721.  \\
%$2s$     &$2s$                     & 72  & 50  & 22.42 & 69  & 665.7 & 72  & 6632.  \\
 $sp$     &$s$                      & 72  & 49  & 2.151 & 68  & 78.31 & 72  & 930.0  \\
 $sp$     &$sp$                     & 144 & 62  & -0.22 & 99  & -22.9 & 128 & 83.44  \\
%$2sp$    &$s$                      & 84  & 52  & 0.238 & 77  & 43.88 & 84  & 513.1  \\
 $spd$    &$s$                      & 132 & 56  & -4.71 & 96  & -8.69 & 123 & 3.119  \\
 $spd$    &$spd$                    & 324 & 71  & -0.66 & 131 & -5.98 & 196 & -20.9  \\
%$2spd$   &$s$                      & 144 & 58  & -15.1 & 99  & -9.44 & 131 & -18.8  \\
%$2s2p2d$ &$s$                      & 240 & 85  & -10.7 & 128 & -41.7 & 194 & -2.18  \\
%$2s2p2d$ &$spd$                    & 432 & 87  & -3.82 & 155 & -5.21 & 243 & -3.87  \\
%$2s2p2d$ &$2s2p2d$                 & 648 & 115 & -1.14 & 193 & -10.1 & 291 & -6.78  \\
\multicolumn{2}{l}{weigend}         & 852 & 86  & -4.39 & 146 & -2.09 & 224 & -2.48  \\
\multicolumn{2}{l}{cc-pvdz-jkfit}   & 1392& 91  & -2.67 & 164 & -1.79 & 253 & -4.43  \\
\hline
\multicolumn{5}{l}{ED threshold $10^{-6}$} \\
 $s$      &$s$                      & 36  & 36  & 117.1 & 36  & 2848. & 36  & 18380. \\
%$2s$     &$s$                      & 48  & 47  & 37.45 & 48  & 1186. & 48  & 9721.  \\
%$2s$     &$2s$                     & 72  & 60  & 16.93 & 72  & 654.3 & 72  & 6632.  \\
 $sp$     &$s$                      & 72  & 59  & 1.313 & 72  & 79.44 & 72  & 930.0  \\
 $sp$     &$sp$                     & 144 & 80  & -0.20 & 117 & 4.703 & 141 & 76.77  \\
%$2sp$    &$s$                      & 84  & 66  & 0.275 & 84  & 36.90 & 84  & 513.1  \\
 $spd$    &$s$                      & 132 & 73  & -0.21 & 115 & -2.67 & 131 & 2.951  \\
 $spd$    &$spd$                    & 324 & 93  & -0.25 & 167 & -1.43 & 235 & -4.75  \\
%$2spd$   &$s$                      & 144 & 76  & -0.72 & 122 & -2.26 & 142 & -1.09  \\
%$2s2p2d$ &$s$                      & 240 & 102 & -4.26 & 163 & -1.13 & 226 & -0.78  \\
%$2s2p2d$ &$spd$                    & 432 & 109 & -0.79 & 202 & -0.29 & 296 & -1.33  \\
%$2s2p2d$ &$2s2p2d$                 & 648 & 142 & -0.35 & 241 & -1.13 & 366 & -0.89  \\
\multicolumn{2}{l}{weigend}         & 852 & 108 & -0.49 & 187 & -0.26 & 291 & -0.16  \\
\multicolumn{2}{l}{cc-pvdz-jkfit}   & 1392& 117 & -0.19 & 209 & -0.22 & 322 & -0.65  \\
\hline
  \end{tabular}
\end{table}

\begin{table}
  \centering
  \caption{The overall errors in HF energy of (H$_2$O)$_{12}$ with PW basis.
  The reference HF energy is -912.40220404.}
  \label{tab:water12:pwbasis}
  \begin{tabular}{llrlrlrlr}
\hline
\multicolumn{2}{l}{PW Basis} & \multicolumn{2}{c}{$\omega=0.1$} & \multicolumn{2}{c}{$\omega=0.15$} & \multicolumn{2}{c}{$\omega=0.2$} \\
$N_{r}$ & $N_\Theta$ & $N_{G}$ & $\Delta E/m E_h$ & $N_{G}$ & $\Delta E/m E_h$ & $N_{G}$ & $\Delta E/m E_h$ \\
%\hline
%15 & 100 &           796  & -10.241 & 796   & -30.157 &  816  &    3.440 \\
%15 & 200 &           1556 &   3.268 & 1568  &  58.613 &  1608 &  278.033 \\
%15 & 300 &           2292 & -0.594  & 2321  & -16.238 &  2377 & -104.032 \\
%15 & 400 &           3168 &   0.005 & 3196  &   2.900 &  3276 &   26.217 \\
%20 & 100 &           944  & -10.222 & 960   & -30.100 &  964  &    3.468 \\
%20 & 200 &           1848 &  3.291  & 1884  &  58.660 &  1896 &  278.054 \\
%20 & 300 &           2726 & -0.575  & 2766  & -16.193 &  2802 & -104.011 \\
%20 & 400 &           3760 &  0.023  & 3828  &   2.946 &  3864 &   26.239 \\
%30 & 400 &           4648 &  0.029  & 4696  &   2.940 &  4708 &   26.229 \\
%\hline
%15 & 100 &           888  & -10.192 & 896   & -30.053 &  896  &    3.494 \\
%15 & 200 &           1724 &   3.320 & 1744  &  58.701 &  1752 &  278.075 \\
%15 & 300 &           2533 &   -.548 & 2573  & -16.152 &  2581 & -103.992 \\
%15 & 400 &           3512 &   0.055 & 3552  &   2.990 &  3568 &   26.262 \\
%20 & 100 &           1016 & -10.195 & 1044  & -30.055 &  1060 &    3.494 \\
%20 & 200 &           2000 &   3.319 & 2044  &  58.702 &  2072 &  278.075 \\
%20 & 300 &           2947 &   -.549 & 3015  & -16.152 &  3043 & -103.992 \\
%20 & 400 &           4076 &   0.054 & 4160  &   2.990 &  4220 &   26.262 \\
%30 & 400 &           5080 &   0.055 & 5108  &   2.990 &  5144 &   26.261 \\
\hline
15 & 100 &           944  & -10.190 & 960   & -30.050 &  964  &    3.496 \\
15 & 200 &           1852 &   3.323 & 1888  &  58.706 &  1896 &  278.077 \\
15 & 300 &           2738 &  -0.546 & 2786  & -16.148 &  2806 & -103.990 \\
15 & 400 &           3772 &   0.057 & 3836  &   2.994 &  3884 &   26.264 \\
20 & 100 &           1096 & -10.190 & 1100  & -30.050 &  1116 &    3.496 \\
20 & 200 &           2144 &   3.323 & 2152  &  58.706 &  2184 &  278.077 \\
20 & 300 &           3159 &  -0.546 & 3176  & -16.148 &  3236 & -103.990 \\
20 & 400 &           4372 &   0.057 & 4388  &   2.994 &  4464 &   26.264 \\
30 & 400 &           5488 &   0.058 & 5508  &   2.994 &  5544 &   26.264 \\
\hline
  \end{tabular}
\end{table}

\begin{table}
  \centering
  \caption{The overall errors in HF energy of Gly$_{6}$ with auxiliary Gaussian basis.
  The reference HF energy is -1316.96782986.}
  \label{tab:gly6:auxbasis}
  \begin{tabular}{llrrlrlrl}
\hline
\multicolumn{2}{l}{Auxiliary Basis} & $N_{aux}$ & \multicolumn{2}{c}{$\omega=0.1$} & \multicolumn{2}{c}{$\omega=0.15$} & \multicolumn{2}{c}{$\omega=0.2$} \\
C,N,O & H & & $N_{ED}$ & $\Delta E/\mu E_h$ & $N_{ED}$ & $\Delta E/\mu E_h$ & $N_{ED}$ & $\Delta E/\mu E_h$ \\
\hline
\multicolumn{5}{l}{ED threshold $10^{-4}$} \\
 $s$      &$s$                    & 45  & 33  &  867.9 & 44  &  7779. & 45  &  35595. \\
%$2s$     &$s$                    & 70  & 40  &  691.5 & 55  &  7339. & 70  &  31613. \\
%$2s$     &$2s$                   & 90  & 45  &  626.7 & 67  &  6717. & 85  &  29479. \\
 $sp$     &$s$                    & 120 & 46  &  20.34 & 71  &  175.5 & 97  &  1224.  \\
 $sp$     &$sp$                   & 180 & 55  &  -3.81 & 96  &  74.34 & 131 &  286.1  \\
%$2sp$    &$s$                    & 145 & 50  &  -21.3 & 81  &  -256. & 116 &  246.3  \\
 $spd$    &$s$                    & 245 & 50  &  14.04 & 87  &  -23.5 & 131 &  -13.9  \\
 $spd$    &$spd$                  & 405 & 59  &  -8.64 & 115 &  -27.4 & 184 &  -56.5  \\
%$2spd$   &$s$                    & 270 & 51  &  -56.1 & 91  &  -226. & 139 &  -58.2  \\
%$2s2p2d$ &$s$                    & 470 & 80  &  -23.0 & 135 &  -364. & 195 &  -99.1  \\
%$2s2p2d$ &$spd$                  & 630 & 83  &  -25.8 & 141 &  -6.43 & 224 &  -102.  \\
%$2s2p2d$ &$2s2p2d$               & 810 & 103 &  -16.4 & 185 &  -58.1 & 272 &  -269.  \\
\hline
\multicolumn{5}{l}{ED threshold $10^{-5}$} \\
 $s$      &$s$                    & 45  & 42  &  751.6 & 45  &  7842. & 45  &  35595. \\
%$2s$     &$s$                    & 70  & 50  &  720.4 & 66  &  7346. & 70  &  31613. \\
%$2s$     &$2s$                   & 90  & 59  &  575.9 & 82  &  6771. & 90  &  29285. \\
 $sp$     &$s$                    & 120 & 60  &  9.802 & 91  &  152.5 & 106 &  1120.  \\
 $sp$     &$sp$                   & 180 & 73  &  -4.04 & 122 &  11.82 & 152 &  149.9  \\
%$2sp$    &$s$                    & 145 & 65  &  -1.07 & 108 &  20.54 & 129 &  274.9  \\
 $spd$    &$s$                    & 245 & 68  &  0.747 & 119 &  -6.92 & 165 &  -51.4  \\
 $spd$    &$spd$                  & 405 & 85  &  0.052 & 156 &  -1.42 & 233 &  -12.4  \\
%$2spd$   &$s$                    & 270 & 70  &  -5.95 & 124 &  -8.51 & 179 &  8.139  \\
%$2s2p2d$ &$s$                    & 470 & 103 &  -18.1 & 172 &  -5.66 & 260 &  -33.6  \\
%$2s2p2d$ &$spd$                  & 630 & 108 &  -17.6 & 191 &  -3.63 & 298 &  -28.1  \\
%$2s2p2d$ &$2s2p2d$               & 810 & 137 &  -0.73 & 231 &  -14.0 & 350 &  -24.0  \\
\hline
\multicolumn{5}{l}{ED threshold $10^{-6}$} \\
 $s$      &$s$                    & 45  & 44  &  744.0 & 45  &  7842. & 45  &  35595. \\
%$2s$     &$s$                    & 70  & 56  &  705.0 & 70  &  7316. & 70  &  31613. \\
%$2s$     &$2s$                   & 90  & 68  &  604.1 & 90  &  6725. & 90  &  29285. \\
 $sp$     &$s$                    & 120 & 70  &  7.205 & 99  &  147.3 & 117 &  1103.  \\
 $sp$     &$sp$                   & 180 & 94  &  0.290 & 139 &  10.05 & 170 &  219.8  \\
%$2sp$    &$s$                    & 145 & 82  &  -0.95 & 120 &  20.43 & 141 &  257.8  \\
 $spd$    &$s$                    & 245 & 90  &  -0.33 & 148 &  -6.35 & 195 &  -49.2  \\
 $spd$    &$spd$                  & 405 & 111 &  -0.23 & 201 &  -0.79 & 280 &  -5.20  \\
%$2spd$   &$s$                    & 270 & 93  &  -2.70 & 158 &  0.794 & 214 &  2.66   \\
%$2s2p2d$ &$s$                    & 470 & 131 &  -0.18 & 221 &  -2.01 & 322 &  -2.05  \\
%$2s2p2d$ &$spd$                  & 630 & 136 &  -0.01 & 244 &  -1.84 & 374 &  -1.42  \\
%$2s2p2d$ &$2s2p2d$               & 810 & 169 &  -0.28 & 289 &  -0.51 & 443 &  -2.10  \\
\hline
  \end{tabular}
\end{table}

%\begin{table}
%  \centering
%  \caption{Errors in HF energy of Gly$_{6}$ with PW basis.
%  The reference HF energy is -1316.96782986.}
%  \label{tab:gly6:pwbasis}
%  \begin{tabular}{llrlrlrlr}
%\hline
%\multicolumn{2}{l}{PW Basis} & \multicolumn{2}{c}{$\omega=0.1$} & \multicolumn{2}{c}{$\omega=0.15$} & \multicolumn{2}{c}{$\omega=0.2$} \\
%$N_{r}$ & $N_\Theta$ & $N_{aux}$ & $\Delta E/E_h$ & $N_{aux}$ & $\Delta E/E_h$ & $N_{aux}$ & $\Delta E/E_h$ \\
%\hline
%20 & 300 & 2726 & 0.00994 & 2766 & 0.079662 & 2802 & -1.31134 \\
%\hline
%20 & 300 & 2947 & 0.00999 & 3015 & 0.079670 & 3043 &          \\
%\hline
%20 & 300 & 3159 & 0.01000 & 3176 & 0.079670 & 3236 & -1.31132 \\
%\hline
%  \end{tabular}
%\end{table}

%\begin{figure*}
%  \includegraphics{image/water128}
%  \caption{SCF convergence rate of water 128 cluster}
%  \label{fig:water128}
%\end{figure*}
%
%\begin{figure*}
%  \includegraphics{image/gly30}
%  \caption{SCF convergence rate of gly-30 molecule}
%  \label{fig:gly30}
%\end{figure*}

\begin{table*}
  \centering
  \caption{Convergence tests for HF with LRDF.
  Convergence threshold for HF energy is set to $10^{-7}$}
  \label{tab:scf}
  \begin{tabular}{lllllcllcll}
\hline
& $N_{AO}$ & \multicolumn{2}{l}{Auxiliary Basis} & $\omega$ & ED thresh & $N_{ED}$ & HF energy/E$_h$ & HF iters. & CPU time/s \\
&        & C,N,O & H & \\
\hline
(H$_2$O)$_{128}$ & 3072 & \multicolumn{5}{c}{conventional analytical ERIs} & -9732.741296 & 11 & 64357 \\
(H$_2$O)$_{128}$ & 3072 & $sp$  & $sp$& 0.1  & $10^{-4}$ & 265 & -9732.741301 & 11 & 34156 \\
(H$_2$O)$_{128}$ & 3072 & $spd$ & $s$ & 0.15 & $10^{-4}$ & 517 & -9732.742271 & 11 & 29299 \\
gly$_{30}$       & 2154 & \multicolumn{5}{c}{conventional analytical ERIs} & -6280.889857 & 13 & 19398\\
gly$_{30}$       & 2154 & $sp$  & $sp$& 0.1  & $10^{-4}$ & 235 & -6280.889816 & 13 & 7479 \\
gly$_{30}$       & 2154 & $spd$ & $s$ & 0.15 & $10^{-4}$ & 382 & -6280.890081 & 13 & 6864 \\
\hline
  \end{tabular}
\end{table*}

\subsection{Auxiliary basis}
To evaluate the impact of auxiliary basis, two groups of auxiliary basis are
considered in these tests:
\begin{itemize}
\item The regular density fitting auxiliary basis (cc-pvdz-jkfit\cite{Weigend2002,Weigend2006}, Weigend
Coulomb-fit basis\cite{Weigend1998});
\item A set of light-weight auxiliary basis.
Each light-weight auxiliary basis only consists of primitive Gaussian orbitals with distinct angular momentum.
The exponents of the Gaussian auxiliary basis are uniformly set to 1.0.
\end{itemize}

In the DF approximation, the main purpose of the auxiliary basis is to describe
the multipoles of the orbital products.
To capture the effects of charge (monopole) on each atom, at least one $s$-type
Gaussian function is required on each atom, which constitutes the smallest basis
set used in the tests.
This smallest basis set represents the worst-case scenario for the LRDF method.
Nevertheless, the error is less than 10 $\mu E_h$ per atom in the
low $\omega$ region (such as $\omega=0.1$).
To achieve a more accurate description of the high order multipoles, polarized
functions such as $p$-type and $d$-type Gaussian functions can be added to the
auxiliary basis.
The inclusion of these functions leads to a significant improvement in accuracy
Even in the worst-case scenario where $\omega=0.2$ and the ED threshold is $10^{-4}$.
The $spd$ basis, which consists of one $s$, one $p$, and one $d$-type Gaussian on
each atom, reduces the error to $0.1 \mu E_h$ per atom for the water
cluster and $1 \mu E_h$ per atom for the linear Gly$_6$ molecule.
When $\omega=0.1$ and ED threshold $10^{-5}$ are utilized, the errors in both
molecules are approximately $0.01\mu E_h$ per atom, which is negligible.

To compare the present LRDF method with the traditional DF method, we also
performed HF energy calculations using the traditional DF method.
The errors of the traditional DF method are significantly larger than the errors of the LRDF
method. For the water cluster, the traditional DF method yields an error of
10003 $\mu E_h$ with the Weigend Coulomb-fit basis and 348 $\mu E_h$ with the
cc-pvdz-jkfit basis. These errors are near the errors of the LRDF method with the smallest auxiliary basis.

Let us now consider the size of the auxiliary basis.
The size of the auxiliary basis is denoted as $N_{aux}$, while the number of linearly independent ED vectors is denoted as $N_{ED}$.
It is worth to note that the regular DF basis is actually over-complete when applied to the LR Coulomb potential metric.
When we employ the ED method to eliminate linear dependency, we find that 70\%
to 95\% of the auxiliary basis can be discarded.
Even in the case of lightweight auxiliary bases, the ED process still reduces the problem size by 70\%.
This suggests that the lightweight auxiliary basis may still be over-complete,
and there are room for further improvement.

\subsection{Range separation parameter $\omega$}
The range separation parameter, $\omega$ controls the exposure of the
local structure of the electron density in the LR DF integrals.
By choosing a small value of $\omega$, fewer short-range details are exposed to
the LR DF part, leading to more accurate results and reduced requirements on the auxiliary basis.
Moreover, a small $\omega$ can effectively reduce the size of linearly independent ED vectors.
As shown in equation \eqref{eq:df:hfx}, the computational cost of DF HFX is
directly proportional to the size of linearly independent ED vectors.
Therefore, a relatively small $\omega$ is favorable due to the $N^4$ scaling in the computation of DF HFX.
However, a smaller $\omega$ results in a slower decay of the SR potential and
increased computational efforts for the analytical SR ERIs.
It is not clear in current study what is the optimal value of $\omega$ that
balances accuracy and computational costs.

\subsection{Threshold of eigenvalue decomposition}
As shown by the $N_{ED}$ columns in the results tables,
the ED threshold has an impact on the size of linearly independent ED vectors, which in
turn affects the computational costs of DF HFX.
In the case of dense systems such as the water cluster, tightening the ED threshold can
significantly improve the accuracy.
In the linear molecule, ED threshold is not the dominant impact compared to the
influence of the auxiliary basis.
With a small value of $\omega$ such as $\omega=0.1$, an ED threshold of $10^{-4}$
is generally sufficient to achieve sub-micro Hartree accuracy.

\subsection{Integration with PW basis}
In Table \ref{tab:water12:pwbasis}, the number of radial grids and the number of
angular grids used for the reciprocal space integral \eqref{eq:lr:gspace}
are represented by the symbols $N_r$ and $N_\Theta$ respectively.
The two parameters are combined to generate all potential plane wave (PW) basis
functions, which are then trimmed based on the resulting integration
weights to reduce computational expenses. The number of remaining PW functions
is denoted as $N_G$ in the table.
  
The accuracy of LR ERIs in this approach relies heavily on the density of integration grids.
As a rough estimation, when $\omega=0.1$, selecting 2 radial grids per Bohr relative to the largest atomic distance in the molecule,
and approximately $N_\Theta \approx N_r^2$ angular grids, can yield an accuracy of approximately $1 \mu E_h$ per atom.
For instance, the water cluster requires $18\times300$ grids, while the linear Gly$_6$ molecule requires $70\times5000$ grids.
After eliminating the PW basis functions that have negligible integration weights,
3000 PWs are still required for the water molecule, while the linear molecule requires 100k PWs.
This is 100 - 1000 times more computationally demanding than the case with Gaussian auxiliary basis.

\subsection{SCF performance}
The SCF convergence achieved with the LRDF method closely matches that of the
SCF with conventional analytical ERIs.
There is no accumulation of errors in larger systems during the SCF iterations.
When $\omega=0.1$, the errors in the
converged HF energy are approximately 5 $\mu E_h$ for the 128 water cluster and
40 $\mu E_h$ for the linear molecule Gly$_{30}$.
The average error is only around $0.1 \mu E_h$ per atom.

The LRDF method significantly reduce computational time.
The water cluster poses a challenge for screening SR integrals since many
atoms are close to each other.
Despite, we observe a reduction in computational time of
50-60\% in the water cluster.
The atoms in the Gly$_{30}$ molecule are more widely distributed.
The integral screening is more effective in this system.
We obtain a 60-70\% reduction in computational time.
It can be expected that the computational cost should be further reduced in larger systems.

The performance of the SR ERI plays a crucial role in this algorithm.
In the Libcint library, ERIs are evaluated with the Rys-quadrature algorithm.
For low angular momentum GTO functions, the quadrature roots and weights of Rys
polynomials are fitted using Chebyshev polynomials when computing the full-range Coulomb integrals.
This polynomial fitting technique is particularly efficient for the full-range ERIs.
However, there are some challenges to develop efficient polynomial fitting schemes for Rys-quadrature in SR ERIs.
In previous versions of Libcint (v5.*), the quadratures were computed from
scratch at runtime, which significantly slows down the performance of SR ERIs.
In Libcint v6.0, SR ERIs are computed as the difference between full-range ERIs
and long-range ERIs for low angular momentum GTO functions.
Although this treatment requires the evaluation of twice the number of integrals,
it is more efficient than calculating the Rys-quadrature of SR ERIs from scratch.
This implies that there is room for improving the algorithm of SR ERIs, which
could further accelerate the SCF performance.

\section{Conclusions}

In this work, we explore the idea of range-separation with long-range density
fitting to compute Coulomb and exchange terms of Hartree-Fock method for molecules.
By combining the analytical algorithm for short-range ERIs and long-range density fitting
ERIs, we find an efficient approach to approximate ERIs for large molecule systems.
Compared to Hartree-Fock with conventional analytical ERIs, the new method achieves more than
two-fold acceleration in performance without compromising accuracy. The error is
around 0.1 $\mu E_h$ per atom.
Our preliminary study shows that the long-range density fitting component has
a lower requirement on the auxiliary basis compared to the regular density fitting method.
Both Gaussian basis and plane-wave basis can be used as the auxiliary basis.
We can achieve errors of 0.1 $\mu E_h$ per atom with a very small auxiliary
Gaussian basis, which is only 1/10 the size of orbital basis.
However, to obtain a similar level of accuracy, it is necessary to use a larger
number of plane-wave basis functions, approximately 100 times more than the
auxiliary Gaussian basis.

\clearpage
\bibliography{lrdf}

\end{document}